\documentstyle[12pt,leqno]{article}

\newtheorem{Lemma}{Lemma}
\newtheorem{Proposition}{Proposition}[section]

\begin{document}

\author{Valeri V.DOLOTIN}
\title{QFT's With Action of Degree 3 and Higher and Degeneracy of Tensors}
\date{May 22, 1997}
\maketitle
\begin{center}
{\it vd@main.mccme.rssi.ru} \\College of Mathematics, Independent 
University of Moscow \\
\end{center}

\begin{abstract}
In this paper we develop a technique of computation of correlation
functions in theories with action being cubic or higher degree form in terms
of discriminants of corresponding tensors. These are analogues of formula $%
\int \exp (iT(x))dx\sim \det T^{-1/2}$ for symmetric tensors of rank two. 
\end{abstract}



\setcounter{section}{-1}
The reader who does not need a motivation may proceed directly with the
technique starting from Section 2.1. In the paper we use terms 
"polynomial", "form" and "tensor" as synonyms, depending on the context.

\section{Introduction}

Let Q be a quadratic form on elements $v$ of a vector space $V$ ({\it %
configuration space}). Then we may take integrals of functions with respect
to the measure of integration $exp(iQ(v))dv$, i.e. take their {\it %
correlation functions}: 
$$
<f>:=\int\limits_Vf(v)\exp (iQ(v))dv 
$$

In particular, when $V$ is n-dimensional and $Q$ is nondegenerate, the
correlation function of constant 1 ({\it partition function}) is given by
the identity:

\begin{enumerate}
\item[(0.1)]  
$$
Z(Q):=<1>\,=\int\limits_V\exp (iQ(x))d^nx=\frac{\pi ^{n/2}}{\left| \det
Q\right| ^{1/2}}\exp \left( \,i\pi \frac{sgnQ}4\right)  
$$
\end{enumerate}

This formula becomes basic if we introduce a parameter k into our measure,
making it as $exp(ikQ(v))dv$ and then let $k\to \infty $. Make a change of
variables $v^{\prime }:=v\sqrt{k}$and write:

$$
<f>=\int\limits_Vf\left( \frac{v^{\prime }}{\sqrt{k}}\right) \exp
(iQ(v^{\prime }))\frac{dv^{^{\prime }}}{k^{n/2}} 
$$

Now as $k\to \infty $ we expand $f$ around 0, make a renormalisation (i.e.
multiply by $k^{n/2}$ in order to get a finite number as a limit) and get

$$
<f>=<f(0)+O\left( \frac 1{\sqrt{k}}\right) >=<1>f(0)+O\left( \frac 1{\sqrt{k}%
}\right) 
$$

So the main impact into correlation function is made by the neighbourhood of
radius $1/\sqrt{k}$ of critical points, where $Q(v)$ with its first
derivatives equals to 0. For nondegenerate $Q$ the only such point is the
origin 0. When $Q$ is degenerate, in the space $V$ besides 0 there is a
distinguished set $C$ of vectors, which are solutions of variational problem
(or stationarity condition):

\begin{enumerate}
\item[(0.2)]  
$$
\delta Q(v)\equiv dQ(v)=0 
$$
\end{enumerate}

This variational problem may be written in the form of homogeneous sytem of
linear equations.

{\bf Example }Let $\dim V=2$. Then 
$$
\delta Q(v)=0\Longleftrightarrow 
\begin{array}{c}
q_{11}v_1+q_{12}v_2=0 \\ 
q_{21}v_1+q_{22}v_2=0 
\end{array}
$$

These are points where $Q(v)$ is ''stationary'' (i.e. with its first
derivatives, Eq.(0.2)) equal to 0, or points of extremum of $Q$. The leading
impact to the integral is made by regions in a neighbourhood of the set $C$
of these critical points. The larger is the value of parameter $k$, the
smaller is the radius of this leading neighbouhood, as we have seen in
nondegenerate case. The integral becomes divergent, but we may make it to
have sense passing to a proper limit and using a proper renormalization
procedure (division by the ''volume of the space'' of solutions $C$). Having
this done, we get for such renormalized integration of a function $f(v)$:

$$
\int\limits_Vf(v)\exp (ikQ(v))dv\sim \frac 1{Vol\
C}\int\limits_Cf(v)dv+O\left( \frac 1{\sqrt{k}}\right) \ 
$$

The higher order ({\it quantum}) terms may be computed again by applying at
each critical point $c$ the basic formular (0.1), but now for the
restriction of $Q$ onto the space transversal to $T_c(C)$. I want to stress
that these considerations are applicable to both finite and infinite
dimensional cases. When the space $V$ is infinite dimensional (for example,
the space of fields $\psi $ on a manifold), the form $Q$ is called {\it %
action} $S(\psi )$ and our integral with a procedure (if there is any) of
assigning a sense to it is called {\it path integral}. The condition, that
the form is degenerate is formulated in terms of existence of $\psi $, which
are solutions for variational problem:%
$$
\delta S(\psi )=0 
$$

This problem is written in the form of {\it field equations} for $\psi $
(the analogue of homogeneous linear system in finite dimensional case). In
infinite dimensional case the solution of these equations is of particular
interest, since using series expansion around these {\it classical states}
is essential for being able to compute correlation functions. So what we
need in order to compute correlation function with the action being
quadratic form - is to compute the determinant of this form and in case when
this form is degenerate - to compute the determinant of its restriction onto
the space, transversal to the set of critical points - solutions of $dQ(v)=0$%
. In infinite dimensional case the procedure of computation of determinant
needs a special definition (zeta-function definition), which at least must
be compatible with finite dimensional case.

\section{Actions of higher order and critical points}

The process of computing the path integral in infinite dimensional case may
be defined as a sequence of finite dimensional integrals and a procedure of
finding a limit of this sequence as the number of variables tends to $\infty 
$. In particular, when the domain of fields $\psi $ is a compact manifold $M$%
, then each triangulation of $M$ gives us a finite set of vertices as the
set of indicies of action form $S(\psi )$. Then we need to compute the
integral with action bieng finite dimensional form and give the procedure of
passing to the limit as the grid of triangulation is getting smaller.

So let us take the action functional as a form on $m$-dimensional vector
space $V$, i.e. a polynomial (in general nonhomogeneous) of $m$ variables $%
P(x_1,...,x_m)$. Take an $m$-cube $R$ in $V$ with center at 0 and side of
length $l$. We may take the integral over $R$ and try to pass to the limit $%
l\rightarrow \infty $:%
$$
Z(P):=\lim _{l\rightarrow \infty }\int\limits_R\exp 
(iP(x_1,...,x_m))dx_1...dx_m $$

\begin{Proposition} The limit $Z(P)$ exists iff the set $C$ of critical
points of function $P(x_1,...,x_m)$ on $V$ is compact.
\end{Proposition}

This set of critical points is given as the solution of variational problem:%
$$
dP(x_1,...,x_m)=0 
$$

We may expect $Z(P)$ to be analitic function of coefficients of $P$, which,
according to Proposition 1.1, has poles at those points which correspond to
forms with noncompact set of critical points.

\section{Homogeneous forms}

Let $T(x)$ be a homogeneous polynomial of $m$ variables $x:=(x_1,...,x_m)$.
Consider the corresponding variational problem:%
$$
dT(x)=0 
$$
In general the number of equations in this system is more then the number of
nonhomogeneous variables, so this system is solvable iff the coefficients of 
$T$ satisfy certain condition.

{\bf Definition }If the variational problem has solutions except $x\equiv
(x_1,...,x_m)=(0,...,0)$ (and, since $T$ is homogeneous, the set of critical
point $C$ is noncompact) the form $T$ is called {\it degenerate. }The set of
degenerate forms is called {\it discriminantal}.

The discriminantal set is described by the equation on coefficients of the
form:%
$$
Dis(T)=0 
$$
where the expression $Dis(T)$ is a polynomial of coefficients of $T$ called
the {\it discriminant}. It is an invariant of the form.

The solvability of variational problem may be reformulated in terms of
geometric properties of the algebraic manifold, corresponding to the
polynomial $T(x)$ as:

\begin{Proposition}
If $Dis(T)=0$ then the algebraic hypersurface $T(x)=0$ 
is singular.
\end{Proposition}

{\bf Example }Let $T(x)=ax_1^2+bx_1x_2+cx_2^2$ be a binary form of degree 2.
The corresponding variational problem is:%
$$
\frac{\partial T(x)}{\partial x_1}=2ax_1+bx_2=0,\ \frac{\partial T(x)}{%
\partial x_2}=bx_1+2cx_2=0 
$$

This system is solvable iff $Dis(T)=\left| 
\begin{array}{cc}
2a & b \\ 
b & 2c 
\end{array}
\right| =b^2-4ac=0$, i.e. the roots of $T(x)=0$ (which are two 1-dimensional
subspaces in the 2-dimensional space of homogeneous variables $(x_1,x_2)$)
merge.

\subsection{$n$-linear forms}

\subsubsection{Basic example}

\begin{Proposition}%
$$
\int\limits_{{\bf R}^1}\int\limits_{{\bf R}^1}\exp ({\it 
i}axy)dxdy=\frac{2\pi }a $$ \end{Proposition}

For a quadratic form $Q$ on $n$-dimensional vector space $V$ we may take its
polarization - a symmetric bilinear form $B(u,v)$ on the pair $U\times V$ of 
$n$-dimensional spaces, such that $Q(v)=B(v,v)$. Let us write the analogue
of formula (0.1) in terms of $B.$ Let $R_1,\;R_2$be two $n$-cubes in $V$ and 
$U$ correspondingly with centers at 0 and sides of length $l$.

\begin{Proposition}
$$
Z(B):=\int\limits_U\int\limits_V\exp (iB(v,u))dvdu\,:=
\lim_{l\rightarrow \infty }
\int\limits_{R_1}\int\limits_{R_2}\exp
(iB(v,u))dvdu=\frac{(2\pi )^n}{\det B} 
$$
\end{Proposition}
So we may understand $\int\limits_V\int\limits_U\exp (iB(v,u))dvdu$ as the
integral taken in this principal value sense.

{\bf Proof }Make a change of variables $v^{\prime }=B^{-1}v,\ u^{\prime }=u.$
Using these variables and Proposition 2.2 we write:%
$$\int\limits_U\int\limits_V\exp 
(iB(v,u))dvdu=\int\limits_U\int\limits_V\exp (i\sum_{k=1}^nv_k^{\prime 
}u_k^{\prime })\frac{dv^{\prime }}{\det B}du^{\prime }=$$ 
$$\frac{1}{\det B}\left( \int \int \exp (ixy)dxdy\right) ^n 
=\frac{(2\pi )^n}{\det B}$$

In bilinear case the condition of degeneracy of $B$, i.e. $d_vB(u,v)=0,\
d_uB(u,v)=0$ is a pair of equivalent linear systems, while each of those is
in turn equivalent to the system $dQ(v)=0$. This equivalence is a specific
of bilinear case.

\subsubsection{General construction}

Let us take as an action functional a $d$-linear form $T$ on the space $%
V_1\oplus \ldots \oplus V_d$, and let $\dim V_i=n$, for each $i$. Having
this form we get the measure of integration $\exp
(iT(v_1,...,v_d))dv_1...dv_d$. Take an $n$-cubes $R_i$ with centers at 0 in
the corresponding $V_i$ and sides of length $l.$ We may take the integral
over $R_1\times \ldots \times R_d$ and try to pass to the limit $%
l\rightarrow \infty $ as in Section 1:

$$
Z(T):=\lim_{l\rightarrow \infty }\int\limits_{R_d}\ldots
\int\limits_{R_1}\exp (iT(v_1,...,v_d))dv_1...dv_d 
$$

To decide whether $C$ is compact we need to write for the form $%
T(v_1,...,v_d)$ (as for all the forms considered so far) the stationarity
conditions:%
$$
\delta T(v)=0\Longleftrightarrow d_{v_1}T(v_1,...,v_d)=0\,,\ \ldots \,,\
d_{v_d}T(v_1,...,v_d)=0 
$$

This is a set of $d$ (one for each $v_i$) nonequivalent (unlike in bilinear
case) systems of $(d-1)$-linear equations.

\subsubsection{Compact set of critical points}

Let the form $T$ be nondegenerate. Then the set of critical points consists
of only one point - the origin. The term ''compact'', applied to this set
will become less redundant in the case of nonhomogeneous forms, considered
below. Let $\deg _T$ denote the degree of {\it $Dis(T).$}

\begin{Proposition}

$$
\int\limits_{V_d}\ldots \int\limits_{V_1}\exp (iT(v_1,...,v_d))dv_1...dv_d:=%
\lim_{l\rightarrow \infty }\int\limits_{R_d}\ldots
\int\limits_{R_1}\exp (iT(v_1,...,v_d))dv_1...dv_d$$

$$=\frac{(2\pi)^{n(d-1)}}{\left| Dis(T)\right| ^{n/\deg _T}}$$ 
\end{Proposition}

Notice, that when $T$ is bilinear $\deg _T=n$, $Dis(T)=\det T$ and we get
the formula (0.1).

So we may understand $Z(T)$ as the integral taken according to this limit
procedure.

{\bf Proof }This is done in the following steps:

\begin{itemize}
\item  for $d=2$ this is the content of Proposition 2.3

\item  $(d+1)$-linear form $T(v_1,...,v_d,v_{d+1})$ is considered as a
linear combination of $d$-linear forms $T^{\prime }(v_1,...,v_d)$, i.e. it
may be viewed as $d$-linear form on $V_1\oplus \ldots \oplus V_d$ with
coefficients $a_{i_1...i_d}^{\prime }$ depending on $v_{d+1}\in V_{d+1}$: 
$$
a_{i_1...i_d}^{\prime
}=\sum\limits_{i_{d+1}}a_{i_1...i_di_{d+1}}(v_{d+1})_{i_{d+1}} 
$$

where $(v_k)_i$ denotes the $i$-th component of vector $v_k\in V_k$.

\item  providing, that the statement is true for $d$-linear forms integrate
over the space $V_1\oplus \ldots \oplus V_d$ to get 
$$
Z(T)=(2\pi )^{n(d-2)}\int\limits_{V_{d+1}}\frac{dv_{d+1}}{|DisT^{\prime
}(v_{d+1})|^{n/\deg _{T^{\prime }}}} 
$$

\item  consider $V_{d+1}$as a real $n_{d+1}$-cycle of integration in the
space ${\bf C}^{n_{d+1}}$ punctured at infinity and compute the last
integral as multidimensional residue of the meromorphic function $%
f(v_{d+1}):=1/DisT^{\prime }(v_{d+1})$.
\end{itemize}

{\bf Example }Let me illustrate this method in the case of $2\times 2\times
2 $ form $T(x,y,z)=\sum\limits_{i,j,k=1,2}a_{ijk}x_iy_jz_k$.

We will use the following fact from complex analysis of many variables:

\begin{Lemma}
For a function $f(z_1,z_2)$ holomorphic in domaine $R\subset 
{\bf C}^2$and $(w_1,w_2)\in R$%
$$
f(z_1,z_2)=\frac 1{(2\pi i)^2}{\int \int }_{|w_i-z_i|=r}\frac{f(w_{1,}w_2)}{%
(w_1-z_1)(w_2-z_2)}dw_1dw_2 
$$
\end{Lemma}
Now for $(x_1,x_2)\in V,\ (y_1,y_2)\in U,\ (z_1,z_2)\in W$ using Proposition
2.3 we have:%
$$
\int\limits_V\int\limits_U\int\limits_W\exp ({\it i}%
\sum_{i,j,k}a_{ijk}x_iy_jz_k)d^2xd^2yd^2z$$
$$=\int\limits_W\left(
\int\limits_V\int\limits_U\exp ({\it i}\sum_{i,j}b_{ij}x_iy_j)d^2xd^2y%
\right) d^2z=(2\pi )^2\int\limits_W\frac{dz_1dz_2}{\det B} 
$$
where $b_{ij}:=\sum\limits_ka_{ijk}z_k$.\\ Then 
$$
\det B=az_1^2+bz_1z_2+cz_2^2 
$$
where $a,b,c$ are expressions of degree 2 of $a_{ijk}.$\\ Make the change of
variables 
$$
\begin{array}{c}
z_1^{\prime }:=((2b+
\sqrt{D})\,z_1+(-2b+\sqrt{D})\,z_2)/2a\sqrt{D} \\ \ z_2^{\prime }:=(z_1-z_2)/%
\sqrt{D}
\end{array}
$$
where $D:=b^2-4ac.$\\ Using these variables we may write 
$$
\int\limits_W\frac{dz_1dz_2}{\det B}=\frac 1{\sqrt{D}}{\int }_W\frac{%
dz_1^{\prime }dz_2^{\prime }}{z_1^{\prime }z_2^{\prime }} 
$$
Considering $W$ as real 2-dimensional cycle in ${\bf C}^2$, the value of the
last integral is computed according to Lemma as 2-dimensional residue of
function $\frac 1{z_1^{\prime }z_2^{\prime }}$ at $0$:%
$$
{\int \int }_{|z_i^{\prime }|=r}\frac{dz_1^{\prime }dz_2^{\prime }}{%
z_1^{\prime }z_2^{\prime }}=(2\pi i)^2 
$$
Substituting these data we get: 
$$
\begin{array}{c}
\int\limits_V\int\limits_U\int\limits_W\exp (
{\it i}\sum_{i,j,k}a_{ijk}x_iy_jz_k)d^2xd^2yd^2z=-\frac{(2\pi )^4}{\sqrt{D}}%
= \\ =-(2\pi
)^4(a_{111}^2a_{222}^2+a_{112}^2a_{221}^2+a_{121}^2a_{212}^2+a_{211}^2a_{122}^2 \\ 
-2a_{111}a_{121}a_{212}a_{222}-2a_{111}a_{211}a_{122}a_{222} \\ 
-2a_{111}a_{112}a_{221}a_{222}-2a_{121}a_{221}a_{112}a_{212} \\ 
-2a_{211}a_{221}a_{112}a_{122}-2a_{212}a_{211}a_{121}a_{122} \\ 
+4a_{111}a_{221}a_{212}a_{122}+4a_{121}a_{211}a_{112}a_{222})^{-1/2}=-\frac{%
(2\pi )^4}{Dis(a_{ijk})^{1/2}}
\end{array}
$$
where $Dis(a_{ijk})$ is the discriminant of 3-linear form, which is:\\

\begin{itemize}
\item  the condition of solvability of the system $d(\sum a_{ijk}x_iy_jz_k)=0
$ of 6 bilinear equations (variational problem)\\

\item  the condition for the set of critical points of function $\sum
a_{ijk}x_iy_jz_k$ on $V\times U\times W$ being noncompact (not descrete)\\

\item  the condition for cubic $\sum a_{ijk}x_iy_jz_k$ being singular.
\end{itemize}

\subsubsection{Noncompact set of critical points}

Let us take as an action functional a $d$-linear form $T$ on the space $%
V_1\oplus \ldots \oplus V_d$ . Having this form we get the measure of
integration $\exp (iT(v_1,...,v_d))dv_1...dv_d$. Let the dimension each $V_i$
be equal to $n_i.$ Take an $n_i$-cubes $R_i$ with centers at 0 in the
corresponding $V_i$ and sides of length $l.$ We may take the integral over $%
R_1\times \ldots \times R_d$ and try to pass to the limit $l\rightarrow
\infty $ as in Section 1:%
$$
Z(T):=\lim_{l\rightarrow \infty }{\lim }\int\limits_{R_d}\ldots
\int\limits_{R_1}\exp (iT(v_1,...,v_d))dv_1...dv_d 
$$
To decide whether $C$ is compact we need to write for the form $%
T(v_1,...,v_d)$ (as for all the forms considered so far) the stationarity
conditions:

\begin{enumerate}
\item[(2.1)]  
$$
\delta T(v)=0\Longleftrightarrow d_{v_1}T(v_1,...,v_d)=0\,,\ \ldots \,,\
d_{v_d}T(v_1,...,v_d)=0 
$$
\end{enumerate}

This is a set of $d$ systems (one for each $v_i$) of $(d-1)$-linear
equations the solutions of which will be critical points for $T.$ As in
bilinear case this homogeneous system is not solvable in general. If we want
this system to have nontrivial (nonzero) solutions, which (since $%
T(v_1,...,v_d)$ is $d$-linear) will form a linear subspace in $V_1\oplus
\ldots \oplus V_d$, then the number of equations must be not more then the
number of nonhomogeneous variables, i.e. we have the condition on
dimensionality of spaces $V_i$: $n_1+...+\stackrel{\wedge }{n_k}%
+...+n_d-d+2<n_k$ for some $k$. In this case the integral diverges. Since in
this case the set $C$ of critical points is a linear subspace in $V_1\oplus
\ldots \oplus V_d$ we may consider the form $T^{\prime }$ induced on the
factor space $V_1\oplus \ldots \oplus V_d/C$, on which the form is
nondegenerate, and so we get into conditions of Section 2.1.3. In terms of
computations, in order to get the finite number as the value of $Z(T)$ we
take the limit:%
$$
Z(T):=\lim_{l\rightarrow \infty }\frac 1{Vol(R_1\times \ldots
\times R_d\cap C)}\int\limits_{R_d}\ldots \int\limits_{R_1}\exp
(iT(v_1,...,v_d))dv_1...dv_d$$
$$=\frac{\Lambda (n_1,...,n_d)}{\left|
Dis(T^{\prime })\right| ^{(n_1+...+n_d)/d\deg _{T^{\prime }}}} 
$$
where $Vol(R_1\times \ldots \times R_d\cap C)$ is computed with respect to
the volume form on $C$ induced form the total space $V_1\oplus \ldots \oplus
V_d$, and $\Lambda (n_1,...,n_d)$ is a constant.

{\bf Example }Let $T=a_1x_1y+a_2x_2y+...+a_nx_ny$ be a bilinear form on $%
V\times U$, where $\dim V=n,\dim U=1$. This form is degenerate, the set of
its critical points is $C=\{(x_1,...,x_n)\in V\ |\
a_1x_1+...+a_nx_n=0\}\times U$, and $\dim C=n-1$.

\begin{Proposition}

$$
Z(T)=\frac 1{Vol(C)}\int \exp (iT(x_1,...,x_n,y))dx_1...dx_ndy=\frac{2\pi }{%
(a_1^4+...+a_n^4)^{1/4}} 
$$
where the division procedure is understood as passing to the limit above.
\end{Proposition}

Notice, that if $n=1$, this formula gives the result of Proposition 2.2.

\subsection{$n$-ary Forms}

Action functionals being considered in physics so far are restrictions of $%
d- $linear forms (on finite or infinite dimensional spaces) onto diagonal of
sets $V_1\oplus \ldots \oplus V_d$, i.e. they are taken as $S(v)=T(v,...,v)$%
. Taking the variational problem we have to differentiate only on one group
of variables $v$ and the resulting equations are not $(d-1)$-linear, but
equations of degree $(d-1)$ on $v$. These equatons may be obtained from any
of $d$ systems of (2.2) by restricting it onto diagonal, i.e. letting $%
v_1=...=v_d=v.$ This restriction causes a symmetry braking in the value of $%
Z(S)$ - it gets a nonzero imaginary part.

\begin{Proposition} 
Let $\dim V=n$. Then%
$$
Z(S):=\int\limits_V\exp (iS(v))dv=\frac{\Lambda (n,d)}{|Dis(S)|^{n/d\deg 
_S}}\exp (i\pi sgn(S)) $$ where $\Lambda (n,d)$ is a constant (which still 
has to be computed for general $n$ and $d$), $Dis(S)$ is the discriminant 
of $S$, i.e. the condition of singularity of the algebraic hypersurface 
$S(v)=0$ in $V$, and the phase $sgn(S)$ is a function on the set of 
connected components of $S^dV^{*}\backslash Dis(S)$.  
\end{Proposition}

The discriminantal manifold $Dis(S)$ makes a partition of the space $S^dV^{*}
$ of $n$-ary forms and the phase function is constant on each of these
component and undergoes a jump when the set of coefficients of $S$ crosses
the discriminantal suface passing to another component. So the ''phase
function'' $sgn(S)$ distiguish the components of the complement to
discriminantal set. The first example of this phenomenon is Formula 0.1.
There $\Lambda =\Lambda (n,2)=\pi ^{n/2}$.

\subsubsection{Binary forms}

Let $\dim V=2$ (case of binary forms, when our physical space consists only
of 2 points), so $S(v)=a_dx^d+ax^{d-1}y+...+a_0y^d.$ In this case $\deg
S=2(d-1)$. Then:

$$
Z(S):=\int\limits_V\exp (i(a_dx^d+ax^{d-1}y+...+a_0y^d))dxdy$$

$$=\frac{\Lambda (2,d)}{|Dis(S)|^{1/d(d-1)}}\exp (i\pi sgn(S))$$ 
here $Dis(S)$ is the 
usual discriminant of polynomial of degree $d$. For binary forms the 
number of component of the complement to discriminantal set has a 
geometric interpretation as the number of real roots of polynomial $S$
the values of $sgn(S)$ on these components are rational numbers, which 
still have to be computed for general case.

{\bf Example} $S=ax^2+bxy+cy^2$. Then $\Lambda (2,2)=\pi $ and the
discriminantal hypersurface partitions 3-dimensional space $S^2V^{*}\ni (a,b,c)$
of coefficients into the following three connected components:%
$$
\begin{array}{c}
D_1=\{\left( \frac b2\right) ^2-ac<0,\ a>0,c>0,\} \\ 
D_2=\{\left( \frac b2\right) ^2-ac<0,\ a<0,c<0\} \\ 
D_3=\{\left( \frac b2\right) ^2-ac>0\} 
\end{array}
$$
Then:%
$$
Z(S):=\int\limits_V\exp (i(ax^2+bxy+cy^2))dxdy=\frac \pi {|\left( \frac 
b2\right) ^2-ac|^{1/2}}\exp (i\pi sgn(S)) $$ where the ''phase function'' 
is:  $$ sgn(S)=\left\{ \begin{array}{c} 1/2,\ \ \ \hbox{for }(a,b,c)\in 
D_1 \\ -1/2,\ \hbox{for }(a,b,c)\in D_2 \\ \ \ \ 0,\ \ \ \ \hbox{for 
}(a,b,c)\in D_3 \end{array} \right\} $$

\section{Nonhomogeneous Forms}

Let the action form $P(v)$ be nonhomogeneous, i.e. the one with terms of
different degrees, for example quadratic and cubic. First notice, that any
nonhomogeneous action may be made homogeneous by introducing one additional
variable and vise versa. The set $C$ of solutions of variational problem $%
\delta P=0$ in general consists of several points, critical points of
function $P$. Notice, that if $v_0$ is a critical point of $P(v)$ then,
making a change of variables $v^{\prime }:=v-v_0$ we have $%
P(v)=c_0+P^{\prime }(v^{\prime })$, where $c_0=P(v_0)$ is constant and $%
P^{\prime }(v^{\prime })$ has terms only of order $\geq 2$. So without loss
of generality we may consider forms without linear terms. According to
Proposition 2.1 we expect $Z(P)$ to be analitic function of coefficients of $%
P$ with poles at those points, where the set of solutions of $\delta P=0$ is
noncompact. Let us call such forms {\it degenerate}, as in homogeneous case.
Notice, that the degeneracy condition on coefficients of $P$ does not
coinside with the degeneracy condition for corresponding homogeneous form
and vise versa. To find $Z(P)$ as an analytic 
function of coefficient of $P$ is an open question in general (even in 
the "initial" case of $P=a_3x^3+a_2x^2$), but let us see an
example, where the analytic properties of $Z(P)$ have a visual 
interpretation in terms of geometric properties of algebraic hypersurface 
$P(v)=0$.

{\bf Example} If we are given a quadratic form $T=\sum_{i,j=0}^nq_{ij}y_iy_j$
on $(n+1)$-dimensional space $V\ni (y_0,y_1,...,y_n)$, then we have a
nonhomogeneous form of $x_i:=\frac{y_i}{y_0}$ as $P=%
\sum_{i,j=1}^na_{ij}x_ix_j+\sum_{i=1}^nb_ix_i+c$, where $%
a_{ij}:=q_{ij},b_i:=a_{i0}+a_{0i}=2a_{i0},c:=a_{00}$. Here in order to give
a more symmetric form to formulas we use summation over all values of $i,j$,
but, of course, $q_{ij}=q_{ji}$.   The variational problem for $P$ is: 
$$
\sum a_{ij}x_j=b_i/2,\ i=1,...,n 
$$
Let $v_0$ be the solution of this nonhomogeneous . Then in terms of $%
v^{\prime }=v-v_0$ we may write $P=\sum_{i,j=1}^na_{ij}x_i^{\prime
}x_j^{\prime }+P(v_0)$. Consider the corresponding partition functions $Z(T)$
and $Z(P)$. The condition of degeneracy of $T$ is $\det (q)=0$, and that for 
$P$ is $\det (a)=0$. So while $Z(T)$ is divergent we still may get a finite
answer for $Z(P)$ integrating over nonhomogeneous variables. And $Z(P)$ gets
in turn divergent just iff the $n\times n$ principal minor of $(q_{ij})$ is $%
0$.

\section{How to compute the discriminant of a form}

In finite dimensional case the algorithm of computing discriminants of $d$%
-linear forms (or determinants of $d$-dimensional matrices) is described in
[1], on computing discriminants of $n$-ary forms see [2]. Of course for
infinite dimensional forms we need a procedure which is the analogue of the
definition of zeta-function for quadratic actions. This will be the subject
of subsequent paper.


\begin{thebibliography}{9}
\bibitem{D1}  V.Dolotin, {\it On Discriminants of Multilinear Forms},
E-print alg-geom/9511010

\bibitem{D2}  V.Dolotin, {\it On Invariant Theory}, E-print 
alg-geom/9512011
\end{thebibliography}
\end{document}